\documentclass[aps,prb,twocolumn,shortbibliography,superscriptaddress]{revtex4-1}
\usepackage{epsfig, epstopdf}
  \usepackage{amsmath, amssymb, amsfonts, bm}
\usepackage{tabularray, longtable} 
\usepackage{makecell, rotating, dcolumn}
\usepackage{hyperref, graphicx}
\usepackage{color, tikz, xcolor}
\usepackage{changes}

\DefTblrTemplate{firsthead, middlehead, lasthead}{default}{}

\begin{document}
\title{Persistent spin textures, altermagnetism and charge-to-spin conversion \linebreak in metallic chiral crystals TM$_{3}$X$_{6}$}

\author{Karma Tenzin}
\affiliation{Zernike Institute for Advanced Materials, University of Groningen, Nijenborgh 3, 9747 AG Groningen, The Netherlands}

\author{Berkay Kilic}
\affiliation{Zernike Institute for Advanced Materials, University of Groningen, Nijenborgh 3, 9747 AG Groningen, The Netherlands}

\author{Raghottam M Sattigeri}
\affiliation{Dipartimento di Fisica, Politecnico di Milano, Piazza Leonardo Da Vinci 32, Milano 20133, Italy}

\author{Zhiren He}
\affiliation{Department of Physics, University of North Texas, Denton, TX 76203, USA}

\author{Chao Chen Ye}
\affiliation{Zernike Institute for Advanced Materials, University of Groningen, Nijenborgh 3, 9747 AG Groningen, The Netherlands}

\author{\linebreak Marcio Costa}
\affiliation{Instituto de F{\'i}sica, Universidade Federal Fluminense, 24210-346, Niter{\'o}i RJ, Brazil}

\author{Marco Buongiorno Nardelli}
\affiliation{Department of Physics, University of North Texas, Denton, TX 76203, USA}
\affiliation{Santa Fe Institute, Santa Fe, NM 87501, USA}

\author{Carmine Autieri}
\email{autieri@magtop.ifpan.edu.pl}
\affiliation{International Research Centre MagTop, Institute of Physics, Polish Academy of Sciences, Aleja Lotnik{\'o}w 32/46, PL-02668 Warsaw, Poland}
\affiliation{SPIN-CNR, UOS Salerno, IT-84084 Fisciano (SA), Italy}

\author{Jagoda S{\l}awi{\'n}ska}
\email{jagoda.slawinska@rug.nl}
\affiliation{Zernike Institute for Advanced Materials, University of Groningen, Nijenborgh 3, 9747 AG Groningen, The Netherlands}

\begin{abstract}
Chiral crystals, due to the lack of inversion and mirror symmetries, exhibit unique spin responses to external fields, enabling physical effects rarely observed in high-symmetry systems. Here, we show that materials from the chiral dichalcogenide family TM$_3$X$_6$ (T = 3d, M = 4d/5d, X = S) exhibit persistent spin texture (PST) -- unidirectional spin polarization of states across large regions of the reciprocal space -- in their nonmagnetic metallic phase. Using the example of NiTa$_{3}$S$_{6}$ and NiNb$_{3}$S$_{6}$, we show that PSTs cover the full Fermi surface, a rare and desirable feature that enables efficient charge-to-spin conversion and suggests long spin lifetimes and coherent spin transport above magnetic ordering temperatures. At low temperatures, the materials that order antiferromagnetically become chiral altermagnets, where spin textures originating from spin-orbit coupling and altermagnetism combine in a way that sensitively depends on the orientation of the N{\'e}el vector. Using symmetry analysis and first-principles calculations, we classify magnetic ground states across the family, identify cases with weak ferromagnetism, and track the evolution of spin textures and charge-to-spin conversion across magnetic phases and different N{\'e}el vector orientations, revealing spin transport signatures that allow one to distinguish N{\'e}el vector directions. These findings establish TM$_3$X$_6$ as a tunable platform for efficient charge-to-spin conversion and spin transport, combining structural chirality, persistent spin textures, and altermagnetism.
 
\end{abstract}

\maketitle

\section{Introduction}
Spin splitting of electronic bands, typically associated with lifted spin degeneracy, originates from two primary symmetry breaking mechanisms. The first is time-reversal ($\mathcal{T}$) symmetry breaking, common in ferromagnets, where exchange interactions induce spin-dependent band structures. The second occurs in materials lacking inversion ($\mathcal{P}$) symmetry, where spin-orbit coupling (SOC) leads to momentum-dependent spin splitting.\cite{zunger_splitting} These band splittings give rise to spin textures - spin polarization patterns of electronic states varying across the Brillouin zone (BZ).\cite{zunger_classification} In particular, the spin polarization at the Fermi surface governs charge-to-spin conversion phenomena and the propagation of spin signals.\cite{Analogs2023} In special cases, crystal symmetries enforce persistent spin textures (PSTs), where spin orientation remains uniform along specific momentum directions,\cite{tsymbal, berkay} enabling unusually long spin lifetimes even in systems with strong SOC.\cite{rondinelli_matter, dou2024spin, xu2024spin} Moreover, PSTs can enhance the charge-to-spin conversion, offering a rare combination of high conversion efficiency and suppressed spin dephasing, two features typically considered mutually exclusive.\cite{evgenii}  

In parallel, a different mechanism for spin splitting has emerged in certain materials with antiferromagnetic coupling, known as altermagnets.\cite{Smejkal2022PRX1, Mazin2022, Hayami2019, Hayami2020} It arises in systems with so-called noninterconvertible spin motif pairs, leading to spin-dependent band structures enforced by magnetic crystal symmetries rather than SOC.\cite{Zunger2023} Altermagnetic spin textures, momentum-dependent spin polarization patterns that often resemble PSTs, enable both conventional and unconventional spin responses. These include the magnetic spin Hall effect (MSHE), a $\mathcal{T}$-odd counterpart of the conventional spin Hall effect,\cite{mshe_original, oppeneer1} as well as the possibility of spin transport with long coherence, particularly in systems where strong SOC is not required.\cite{Spinsplitter2021PRL} By combining features of ferromagnets and antiferromagnets, altermagnets have sparked broad interest across spintronics, spin caloritronics,\cite{Zhou2023PRL} superconductivity,\cite{Ouassou2023PRL} and related technologies.\cite{bai2024altermagnetism,tamang2025altermagnetism,Song2025}

To date, altermagnets have commonly been studied in centrosymmetric crystals, and their realization in noncentrosymmetric crystals offers a still largely unexplored opportunity.\cite{cheong2025altermagnetism} In this context, chiral materials, lacking both inversion and mirror symmetries, are particularly promising, as they naturally support altermagnetism upon antiferromagnetic ordering. 
Moreover, the reduced symmetry of chiral systems also favors the emergence of robust PSTs in their nonmagnetic phases.\cite{berkay} The coexistence of PSTs and altermagnetism raises compelling questions: How do PSTs evolve upon antiferromagnetic ordering? Can spin textures driven by SOC and those originating from altermagnetism interfere or reinforce one another? What are the implications for charge-to-spin conversion and spin transport? While recent work has examined the Rashba-Edelstein effect (REE) in noncollinear altermagnets,\cite{hu2024spin} the interplay between PSTs and altermagnetism in chiral systems remains unexplored.

In this paper, we explore the chiral dichalcogenide family NiX$_3$S$_6$ (X = Ta, Nb), which combines strong SOC with chirality and altermagnetism. In the nonmagnetic metallic phase, we identify symmetry-enforced PSTs that span over entire Fermi surfaces and drive very efficient charge-to-spin conversion. Upon entering the low-temperature phases, these systems exhibit rich magnetic behavior, including altermagnetism and weak ferromagnetism, governed by the symmetry of the magnetic order. We analyze how these magnetic ground states reshape spin textures and charge-to-spin conversion using first-principles calculations complemented by symmetry analysis. To comprehensively address charge-to-spin conversion in both nonmagnetic and magnetic phases, we implement spin Hall effect (SHE) and Rashba-Edelstein effect, including the $\mathcal{T}$-even and $\mathcal{T}$-odd variants, within the open-source \textsc{paoflow} package, extending its capabilities to magnetic systems and laying the groundwork for high-throughput discovery of spintronic materials.

\section{Results and discussion}
\subsection{Calculation of charge-to-spin conversion}
To study charge-to-spin conversion in nonmagnetic and magnetic materials, we perform ground-state electronic structure calculations via density functional theory (DFT) as implemented in Vienna Ab initio Simulation Package (\textsc{vasp}),\cite{vasp1, vasp2, vasp3} and use the obtained \textit{ab initio} wave functions to generate PAO Hamiltonians in the \textsc{paoflow} code.\cite{PAOFLOW1, PAOFLOW2} Our implementation of the spin Hall effects and Rashba-Edelstein effects is based on the Kubo linear response formalism.\cite{manchon, freimuth} In the constant relaxation time approximation, an observable $\delta \mathbf{A}$ induced in response to an external electric field \textbf{E} is expressed as $\delta A_i = (\chi ^I_{ij} + \chi^{II}_{ij}) E_j$, where

\small
\begin{eqnarray}
\chi_{i j}^I=-\frac{e \hbar}{\pi} \sum_{\mathbf{k}, n,m} \frac{\Gamma ^2\operatorname{Re}\left[\left\langle\psi_{\mathbf{k} n}\right| \hat{A}_i\left|\psi_{\mathbf{k} m}\right\rangle\left\langle\psi_{\mathbf{k} m}\right| \hat{v}_j\left|\psi_{\mathbf{k} n}\right\rangle\right]}{\left(\left(E_F-E_{\mathbf{k} n}\right)^2+\Gamma^2\right)\left(\left(E_F-E_{\mathbf{k} m}\right)^2+\Gamma^2\right)}\nonumber\\
\end{eqnarray}


\begin{eqnarray}
\begin{aligned}
\chi_{ij}^{II} = 2e\hbar &\sum^{\substack{\textit{\tiny{n~occ.}} \\ \textit{\tiny{m~unocc.}}}}_{\mathbf{k},n\neq m} \frac{\operatorname{Im}\left[\left\langle\psi_{\mathbf{k}n}\right| \hat{A}_i\left|\psi_{\mathbf{k} m}\right\rangle\left\langle\psi_{\mathbf{k}m}\right|\hat{v}_j\left|\psi_{\mathbf{k}n}\right\rangle\right]}{\left[ \left(E_{\mathbf{k}n}-E_{\mathbf{k}m}\right)^2+\Gamma^2\right]^2} \\ \\
&\hspace{+8.5ex}\times\left(\Gamma^2- (E_{\mathbf{k}n}-E_{\mathbf{k}m})^2\right)
\end{aligned}
\end{eqnarray}

\normalsize

\noindent In the above equations, $e$ is the elementary (positive) charge, \textbf{k} is the Bloch wave vector, $n, m$ are the band numbers,  $E_{kn}$ is the band energy, $E_F$ is the Fermi energy, $\hat{v}$ is the velocity operator, and $\Gamma$ is a disorder parameter related to the relaxation time $\tau$ as $\Gamma=\frac{\hbar}{2 \tau}$. 

The separation of the response into terms (1) and (2) is motivated by their distinct transformation under the time-reversal symmetry operation, and whether they are $\mathcal{T}$-even or $\mathcal{T}$-odd depends on the choice of the operator $\hat{A}$. When $\hat{A}$ is a spin operator, $\chi^{I}$ is $\mathcal{T}$-even and corresponds to the standard Rashba-Edelstein effect. Note that assuming weak scattering, it can be viewed as a sum of two terms: (i) the Fermi surface (intraband) term for $n = m$ case, and (b) the Fermi sea (interband) term for $n \neq m$ case, which vanishes in the limit of $\Gamma \rightarrow 0$.\cite{manchon} The term $\chi^{II}$ describes the $\mathcal{T}$-odd Rashba-Edelstein effect recently studied in noncollinear magnetic systems.\cite{GH2024} In the present study, we assume low $\Gamma$, and the terms in Eq. (2) that are quadratic in $\Gamma$ are neglected in the calculations. 

In contrast, when $\hat{A}$ represents the spin current operator $\hat{J}^i_{l} = \frac{1}{2}\left\{\hat{s}_i, \hat{v}_l\right\}$, $\chi^{I}$ is $\mathcal{T}$-odd and describes magnetic spin Hall effects previously studied in ferro- and antiferromagnets.\cite{oppeneer1, zelezny_noncollinear_afm} The expression for $\chi^{II}$ is then $\mathcal{T}$-even and corresponds to the intrinsic spin Hall conductivity commonly studied in nonmagnetic materials with strong SOC. Following previous works, we assume that $\Gamma$ is low, and $\chi^{II}$ reduces to a well-known spin Hall conductivity (SHC) expression:

\begin{equation}
\chi_{i j}^{I I} \approx-2 e \hbar \sum^{\substack{\textit{\tiny{n~occ.}} \\ \textit{\tiny{m~unocc.}}}}_{\mathbf{k},n\neq m} \frac{\operatorname{Im}\left[\left\langle\psi_{k n}\right| \hat{J}^i_l\left|\psi_{k m}\right\rangle\left\langle\psi_{k m}\right| \hat{v}_j\left|\psi_{k n}\right\rangle\right]}{\left(E_{k n}-E_{k m}\right)^2}
\label{final_chi_ii}
\end{equation}


\subsection{Chiral magnetic dichalcogenides TM$_3$X$_6$}
The chiral dichalcogenides TM$_3$X$_6$ (T = 3d transition metal, such as Ni, Cr, Mn, Co, V or Fe; M = Nb or Ta; X = S or Se) have been studied extensively over the past few decades.\cite{parkin_family} They crystallize in the chiral space group P6$_3$22 (No. 182) and consist of hexagonal dichalcogenide layers (2H-MX$_2$) intercalated with 3d transition metal atoms, as illustrated in Fig. 1a. The intercalation leads to the formation of $(\sqrt{3} \times \sqrt{3})R30^\circ$ superstructure relative to the $(1\times1)$ unit cell of the parent dichalcogenide. The TM$_3$X$_6$ family reveals a diverse range of magnetic phases, including ferromagnetism, helimagnetism, and collinear antiferromagnetism.\cite{Boucher2024, CoNbS_weak_ferro, CoxNbS2, CoTaS, MnNb3S6, Kousaka2022} While the magnetic structure is typically linked to the choice of 3d transition metal (e.g., Ni and Co favor antiferromagnetic coupling, while Cr often leads to ferromagnetism), several exceptions are known, such as the helimagnetic CrNb$_3$S$_6$.\cite{Kousaka2022} Also, recent studies have examined Ni-based systems NiTa$_3$S$_6$ and NiNb$_3$S$_6$, revealing different magnetic behaviors: NiTa$_3$S$_6$ is a collinear antiferromagnet, whereas NiNb$_3$S$_6$ exhibits a helimagnetic order.\cite{An2023PRB}

Surprisingly, altermagnetism in TM$_3$X$_6$ materials has been explored in detail only recently.\cite{Ray2025} Structural chirality is defined by the absence of mirror planes and inversion centers, with symmetries restricted to rotations and translations. These rotational symmetries can connect spin-up and spin-down sublattices, supporting noninterconvertible spin-structure motifs. In TM$_3$X$_6$ antiferromagnets, the two Ni sites are not connected by either mirror, inversion, or translation symmetries, which allows the presence of altermagnetic splitting. Differently to centrosymmetric materials, TM$_3$X$_6$ will additionally have a SOC-induced band splitting leading to an interesting interplay of the two effects, which was so far unexplored. In the following sections, we present a general symmetry analysis for altermagnetic TM$_3$X$_6$, and discuss the electronic structures and charge-to-spin conversion phenomena based on DFT calculations performed for representative materials NiTa$_3$S$_6$ and NiNb$_3$S$_6$.

\begin{figure}    
\centering
\includegraphics[width=1\linewidth]{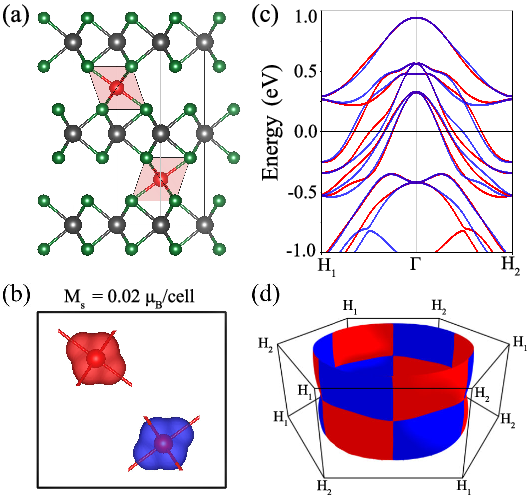}
\caption{Geometry and nonrelativistic electronic properties of altermagnetic NiTa$_3$S$_6$. (a) Crystal structure. (b) Magnetization density isosurface ($M_s= \pm $0.02 $ \mu_B$/cell) calculated around the Ni sites. (c) Band structure along the $H_1-\Gamma-H_2$ line. (d) Spin-polarized Fermi surface ($E=E_F$); only the outermost pair of Fermi sheets is shown. The red and blue colors correspond to spin-up and spin-down states, respectively. 
}
\end{figure}

\begin{figure*}[ht!]
\centering
\includegraphics[width=1\textwidth]{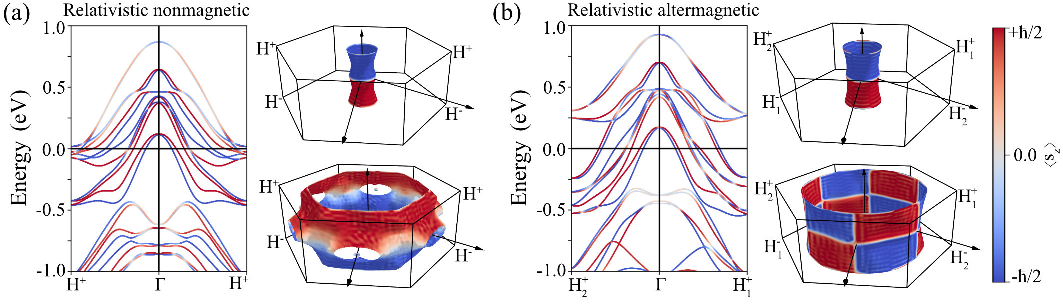}
\caption{Relativistic electronic structure of NiTa$_3$S$_6$. (a) Band structure calculated for the nonmagnetic phase (left-hand panel), and the representative Fermi sheets (right-hand panels). The narrow cylindrical sheet corresponds to the innermost band, while the larger sheet below originates from one of the outermost bands. Note that the inequivalent high-symmetry points $H^{+}$ and $H^{-}$ are different from those in the nonrelativistic magnetic phase shown in Fig. 1. (b) Same as (a) but calculated for the altermagnetic phase with the N{\'e}el vector along the [001] direction. The color of the bands represents the $S_z$ projection of spin texture; $S_x$ and $S_y$ are negligible and they are not shown. Note that the inequivalent high-symmetry points are different from those in Fig. 1. The full set of Fermi surfaces for both phases is provided in the Supplementary Information.}
\end{figure*}

\subsection{Symmetry, weak ferromagnetism and altermagnetism}
We start with the general analysis of magnetism in collinear antiferromagnets TM$_3$X$_6$, assuming the magnetic unit cell equivalent to the crystal primitive cell (see Fig.~1a). We focus on different directions of N{\'e}el vector observed in different materials and the potential presence of weak ferromagnetism. To this aim, we apply the Landau theory of altermagnetism formulated by McClarty and Rau.\cite{Landautheory2024PRL} Weak ferromagnetism in its simplest form microscopically originates from the staggered Dzyaloshinskii–Moriya (DMI) interaction,\cite{Dzyaloshinsky1958, Autieri2025} and it is not allowed in the nonrelativistic limit. The Landau theory of phase transitions connects the zero SOC altermagnetic limit to the finite SOC phase, where weak ferromagnetism with magnetization $\mathbf{M}$ arises as a secondary order parameter coupled to the staggered magnetization represented by a N{\'e}el vector $\mathbf{N}$. 

In linear order, such terms can appear in the order parameter as $\propto \mathbf{N}\cdot\mathbf{M}$; and whether they are allowed or not is fully determined by the point group of the magnetic system, which depends on the crystal structure as well as the positions of the magnetic ions and the magnetization direction. Specifically, such a coupling term is allowed when it is invariant under the symmetries of the point group.
If the N{\'e}el vector is along the $z$-axis, as in the ground state of NiTa$_3$S$_6$,\cite{An2023PRB} the magnetic structure transforms under the point group $D_6$. For the N{\'e}el vector on the $xy$-plane, which approximately holds for NiNb$_3$S$_6$,\cite{An2023PRB} the corresponding point group is $D_2$. The transformation properties of $\mathbf{N}$ and $\mathbf{M}$ can be understood using irreducible representations (irreps) of these point groups.

Let us consider first the limit of zero SOC. As axial vectors, $\mathbf{M}$ transforms as $\Gamma_1 \otimes \Gamma^S_A$ and $\mathbf{N}$ transforms as $\Gamma_N \otimes \Gamma^S_A$ where $\Gamma_1$ and $\Gamma_N$ are the trivial and nontrivial irrep of the point group, and $\Gamma^S_A$ is the axial vector irrep of the spin point group.\cite{mcclartySPG} $\Gamma_N$ being nontrivial is a signature of altermagnets. A linear coupling term $\mathbf{N}\cdot\mathbf{M}$ thus transforms as $(\Gamma_1 \otimes \Gamma^S_A) \otimes (\Gamma_N \otimes \Gamma^S_A) = \Gamma_N \otimes (\Gamma^S_1 \oplus \Gamma^S_A \oplus \Gamma^S_Q)$ where $\Gamma^S_1$ and $\Gamma^S_Q$ are the trivial and quadrupole irreps of the spin point group. Since $\Gamma_N$ is nontrivial, this term is not invariant under the point group symmetries and weak ferromagnetism is not allowed in the absence of SOC.

When SOC is turned on, the spin degrees of freedom lock onto the crystal symmetries; thus, the spin point group is reduced to the crystallographic point group: $\Gamma^S_\alpha \rightarrow \Gamma_\alpha$, where $\alpha = 1,A,Q$ is used to denote the corresponding irrep. $\mathbf{N}\cdot\mathbf{M}$ thus transforms as $(\Gamma_1 \otimes \Gamma_A) \otimes (\Gamma_N \otimes \Gamma_A) = \Gamma_N \otimes (\Gamma_A \otimes \Gamma_A)$. Using the property $\Gamma_A \otimes \Gamma_A = \Gamma_1 \oplus \Gamma_A \oplus \Gamma_Q$, we see that linear coupling is only allowed when $\Gamma_N \subseteq \Gamma_A$ or $\Gamma_Q$. For both NiTa$_3$S$_6$ and NiNb$_3$S$_6$ the point group is $D_6$, $\Gamma_N = B_2$, $\Gamma_A = A_2 \oplus E_1$, and $\Gamma_Q = A_1 \oplus E_1 \oplus E_2$. Since $B_2$ is contained neither in $\Gamma_A$ nor $\Gamma_Q$, weak ferromagnetism in linear order is not allowed in any of the two materials. 


Next, we will consider higher order couplings $\mathbf{N}^n\cdot\mathbf{M}$.
In NiTa$_3$S$_6$, the out-of plane N{\'e}el vector $\mathbf{N}=N\mathbf{\hat{z}}$ spatially transforms as $\Gamma_A = A_2$, and in-plane magnetization $\mathbf{M}=M_x\mathbf{\hat{x}}+M_y\mathbf{\hat{y}}$ transforms as $E_1$. The higher order coupling term thus transforms as $(B_2 \otimes A_2)^n \otimes E_1$ = $B_1^n \otimes E_1$ are disjoint for all $n$. Thus, even higher-order weak ferromagnetism is not allowed in NiTa$_3$S$_6$. On the other hand, for an in-plane N{\'e}el vector $\mathbf{N} = N_x\mathbf{\hat{x}}+N_y\mathbf{\hat{y}}$ and out-of-plane magnetization $\mathbf{M}||\hat{z}$, the higher order coupling term transforms as $(B_2 \otimes E_1)^n \otimes A_2$, which is allowed for odd values of $n$. For $n=3$, this term is $N_x(N_x^2-3N_y^2)M_z$. Therefore, weak ferromagnetism in NiNb$_3$S$_6$ is allowed by symmetry only at the third or higher order in the free-energy expansion.

It is worthwhile to note that the representations of the point groups are also helpful for understanding the symmetry of the nonrelativistic spin splitting of bands. The lowest order multipole that is allowed to linearly couple to $\mathbf{N}$ determines the type of altermagnetic splitting. For point group $D_6$ in the absence of spin-orbit coupling, it is the $l=4$ (g-wave) multipole, and the part that transforms as $B_1$ is $k_yk_z(3k_x^2 - k_y^2)$, reflecting the g-wave spin splitting pattern of Fermi surfaces predicted previously.\cite{Smejkal2022PRX1} 


\subsection{Persistent spin textures in nonmagnetic and altermagnetic phases
}
Let us consider first the ideal nonrelativistic altermagnetic case. The results of our nonrelativistic DFT calculations performed for NiTa$_3$S$_6$ are summarized in Fig. 1. The material is metallic and exhibits a large Fermi surface, consisting of four pairs of sheets that differ in size, with the largest shown in Fig. 1d. The spin polarization exhibits a g-wave character, consistent with the symmetry analysis. Notably, band structures calculated along the high-symmetry lines will not show any altermagnetic splitting. To illustrate the splitting along the $k$-lines, we chose a diagonal path $H_1-\Gamma-H_2$, which reveals a maximal altermagnetic splitting of bands, reaching up to 100 meV near the Fermi level (see Fig. 1c). Note that the material hosts opposite spin-splitting for
 $\Gamma-H_1$ and $\Gamma-H_2$ paths, consistently with the g-wave pattern.

To explore the impact of SOC, we performed fully relativistic calculations for both nonmagnetic and antiferromagnetic phases, considering different orientations of the N{\'e}el vector. For NiTa$_3$S$_6$, the magnetic anisotropy energy (MAE) calculations indicate that the $z$-axis alignment of the  N{\'e}el vector is favored over the $x$ and $y$, 
in agreement with the previous studies.\cite{An2023PRB} Additionally, this magnetic configuration has not revealed any signs of weak ferromagnetism, consistent with the group theory analysis presented in Sec. IIC. The relativistic electronic structures of both nonmagnetic and altermagnetic phases are displayed in Fig. 2a and Fig 2b, respectively. Although the band structures and spin textures appear similar, the fundamental symmetry-related differences can lead to distinct behavior in response to the external stimuli. 

In the nonmagnetic phase, the spin splitting of bands and the spin texture arise solely from SOC. Figure 2a displays the dominant $S_z$ component of the spin texture, while the $S_x$ and $S_y$ components remain negligible for most bands (see Fig. S1-S3 in the Supplementary Information). This uniform spin polarization, called a persistent spin texture, is enforced by the crystal symmetries.\cite{tsymbal, berkay} Following the analysis from Ref. \onlinecite{berkay}, we notice that SG 182 enforces PST around the high-symmetry points $\Gamma$ and $H$ for \textit{some} of the bands. This is clearly observed in Fig. 2a: most states exhibit PST, except for the bands with onsets at 0.75 eV and -0.5 eV, which lack PST around $\Gamma$ and $H$ or only at $\Gamma$, respectively. This behavior is dictated by the representations of symmetry operations at these $k$-points, which impose distinct constraints on the spin expectation values for different bands. Notably, all large Fermi surfaces ($E = E_F$) exhibit PST throughout their entire extent, which is surprising for large-sized sheets in a metallic bulk material. Similar to the case of chiral tellurium, this uniform symmetry-enforced spin texture is expected to enable a strong Rashba-Edelstein effect and potentially support long spin lifetimes.\cite{evgenii}

These observations raise an important question: does the relativistic altermagnetic electronic structure inherit the nonrelativistic altermagnetic spin splitting, the nonmagnetic PST, or a combination of both? The results shown in Fig. 2b and Fig. S4-S5 suggest a band-dependent behavior. Among the bands forming the Fermi surface, the outermost pair exhibits an altermagnetic spin polarization pattern, while the rest retain PST. This can be attributed to different orbital character, with the outer Fermi surfaces dominated by Ni states and the inner bands showing stronger Ta–S contributions. Since both textures involve out-of-plane ($S_z$) spin components, their coexistence should be, in principle, constructive, potentially enhancing charge-to-spin conversion and spin coherence. Interestingly, the inclusion of magnetic order appears to reduce the overall band splitting.

The electronic structure of the nonmagnetic phase of NiNb$_3$S$_6$ follows similar trends of NiTa$_3$S$_6$. All Fermi surfaces exhibit fully persistent spin textures along the $z$ axis (Figs. S6–S7). However, the magnetic configuration reveals quite different properties. First of all, the nonrelativistic altermagnetic band splitting is negligible (Fig. S13a). While it can be enhanced by introducing a Hubbard $U$ term, leading to a splitting that can even exceed that of NiTa$_3$S$_6$ (see Fig. S13b), the physical justification of such an approach for a metallic system is subtle, and the results should be interpreted cautiously in light of experimental evidence. The relativistic calculations with the N{\'e}el vector aligned along the $x$-axis show that, although the spin textures still reflect an interplay between PST and altermagnetism, their character differs notably from those in NiTa$_3$S$_6$ (see Figs. S8–S10). Specifically, the two inner pairs of Fermi surfaces retain persistent spin textures dominated by the $S_z$ component, accompanied by a g-wave-like $S_x$ pattern. In contrast, the outer Fermi surfaces are almost entirely altermagnetic, exhibiting strong $S_x$-polarized spin textures. This hybrid configuration, with competing $S_x$ and $S_z$ components, reflects a complex relativistic spin structure that cannot be captured by a nonrelativistic approximation. Additionally, NiNb$_3$S$_6$ displays weak ferromagnetism and an anomalous Hall effect, in agreement with our symmetry predictions (see Sec. S4, Fig. S14).

\subsection{Collinear Rashba-Edelstein effect}



We now examine charge-to-spin conversion mechanisms in both the nonmagnetic and altermagnetic phases, beginning with the Rashba-Edelstein effect. This phenomenon refers to spin accumulation induced by a charge current, where the relative orientation of the current and generated spin density is dictated by crystal symmetry.\cite{Analogs2023} In the nonmagnetic phase, only the $\mathcal{T}$-even component of the REE is allowed. The symmetry analysis of the response tensor further shows that only the so-called collinear components of REE will be present, meaning that the induced spin polarization aligns with the direction of the charge current, consistent with typical behavior in chiral materials.\cite{ Roy2022npj, Tenzin2023PRB,  slawinska2023spin} In contrast, in the altermagnetic phase, the response depends on the orientation of the N{\'e}el vector. For NiTa$_3$S$_6$, with the N{\'e}el vector along the $z$-axis, the symmetry again allows only $\mathcal{T}$-even components, while $\mathcal{T}$-odd terms are forbidden. However, in NiNb$_3$S$_6$, where the N{\'e}el vector lies in the plane, several $\mathcal{T}$-odd components emerge in addition to the $\mathcal{T}$-even diagonal elements. These components, marked in red and blue in Table I, suggest a much richer REE response. 

\begin{figure}
    \centering
    \includegraphics[width=\linewidth]{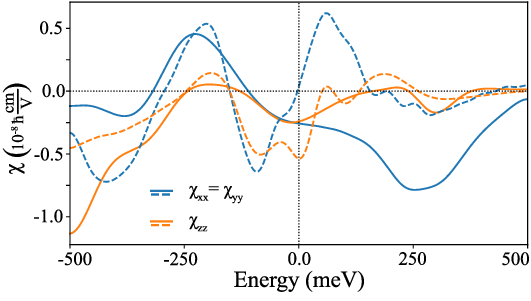}
    \caption{ Collinear Rashba-Edelstein effect in NiTa$_3$S$_6$. $\mathcal{T}$-even REE response tensor $\chi$ vs chemical potential calculated for nonmagnetic phase (dashed lines), and altermagnetic phase with the N\'eel vector along the \textit{z-}direction (solid lines). In both cases, we used the parameter $\Gamma$ = 0.03 eV determined from the comparison of measured and calculated charge conductivity (see Sec. S3 in the Supplementary Information.) 
    }
    \label{fig:rel_ree}
\end{figure}

Our calculations for NiTa$_3$S$_6$ confirm the presence of three $\mathcal{T}$-even components of the Rashba-Edelstein tensor. The calculated values of $\chi$ as a function of chemical potential are shown in Fig. \ref{fig:rel_ree}, where dashed and solid lines represent the nonmagnetic and altermagnetic calculations, respectively. In the considered energy range, the maximal magnitudes of $\chi$ are comparable between the two phases and remain large, on par with or exceeding those of other chiral materials such as TaSi$_2$ and NbSi$_2$ (see Fig. S18),\cite{Roy2022npj, Tenzin2023PRB} or previously studied noncollinear antiferromagnets.\cite{GH2024} At the true Fermi level in the nonmagnetic phase, the values of $\chi_{zz}$ significantly exceed $\chi_{xx}=\chi_{yy}$ which are nearly zero, consistent with the almost full spin polarization of states along the $z$ axis. Such a strong response highlights the potential of this material for charge-to-spin conversion, where the presence of PST appears to play a major role in enhancing the REE. 


The REE response of NiNb$_3$S$_6$ presented in Fig. S15 reveals three notable features that distinguish it from NiTa$_3$S$_6$. First, the $\mathcal{T}$-even response in the nonmagnetic phase is slightly larger and exhibits an opposite sign at the Fermi level compared to NiTa$_3$S$_6$, highlighting subtle differences in the underlying spin textures. Second, in the altermagnetic phase, a pronounced anisotropy emerges between $\chi_{xx}$, $\chi_{yy}$, and $\chi_{zz}$ components, accompanied by an overall suppression of the $\mathcal{T}$-even REE. This reduction arises from the competing contributions of $S_x$ and $S_z$ spin texture components on the Fermi surface, reflecting the complex interplay between PST and altermagnetism. Third, several $\mathcal{T}$-odd components appear in NiNb$_3$S$_6$, which are symmetry-forbidden in NiTa$_3$S$_6$, but their magnitude remains about an order of magnitude smaller than the $\mathcal{T}$-even terms in the nonmagnetic phase. This richer charge-to-spin conversion response, particularly the presence of $\mathcal{T}$-odd components, offers a clear fingerprint of the N{\'e}el vector orientation and could serve as an experimental probe to detect and distinguish antiferromagnetic order in this material class.

\subsection{Spin Hall effect in nonmagnetic and altermagnetic phases}

\begin{table*}
\renewcommand{\arraystretch}{1.5} 
\caption{Magnetic point group and allowed response tensors for the different N{\'e}el vectors $\mathbf{N}$ directions. $\mathcal{T}$-even and $\mathcal{T}$-odd components are highlighted in black and red colors, respectively. Components marked in blue allow both $\mathcal{T}$-even and $\mathcal{T}$-odd.}
\begin{longtblr}[caption = {!!!THIS IS DELETED!!!}, label = {shc-table}]{colspec = {c|c|X|X|X|X}}
\hline
\hline
\SetCell[r=2]{c}{\centering{Néel vector}} & \SetCell[r=2]{c}{\centering{MPG}} & \SetCell[c=3]{c}{SHC} & & &
\SetCell[r=2]{c}{\centering{REE}} \\* \hline
 & & \centering{$\chi^{x}$} & \centering{$\chi^{y}$} & \centering{$\chi^{z}$} & \\* \hline
\centering{$\mathbf{N}\parallel [001]$} & \centering{$6^{\prime}22^{\prime}$} &
\centering{$\begin{pmatrix} 
\textcolor{red}{\sigma^{x}_{xx}} & 0 & 0 \\ 
0 & \textcolor{red}{-\sigma^{x}_{xx}} & -\sigma^{y}_{xz} \\ 
0 & -\sigma^{z}_{xy} & 0
\end{pmatrix}$} 
&  
\centering{$\begin{pmatrix} 
0 & \textcolor{red}{-\sigma^{x}_{xx}} & \sigma^{y}_{xz} \\ 
\textcolor{red}{-\sigma^{x}_{xx}} & 0 & 0 \\ 
\sigma^{z}_{xy} & 0 & 0
\end{pmatrix}$} 
& 
\centering{$\begin{pmatrix} 
0 & \sigma^{y}_{zx} & 0 \\ 
-\sigma^{y}_{zx} & 0 & 0 \\ 
0 & 0 & 0
\end{pmatrix}$}
&

\centering{$\begin{pmatrix} 
\sigma_{xx} & 0 & 0 \\ 
0 & \sigma_{xx} & 0 \\ 
0 & 0 & \sigma_{zz}
\end{pmatrix}$}

\\*
\hline
\centering{$\mathbf{N} \parallel [100]$} & \centering{$2^{\prime}2^{\prime}2$} &
\centering{$\begin{pmatrix} 
0 & 0 & \textcolor{red}{\sigma^{x}_{xz}} \\ 
0 & 0 & \sigma^{x}_{yz} \\ 
\textcolor{red}{\sigma^{x}_{zx}} & \sigma^{x}_{zy} & 0
\end{pmatrix}$} 
&
\centering{$\begin{pmatrix} 
0 & 0 & \sigma^{y}_{xz} \\ 
0 & 0 & \textcolor{red}{\sigma^{y}_{yz}} \\ 
\sigma^{y}_{zx} & \textcolor{red}{\sigma^{y}_{zy}} & 0
\end{pmatrix}$} 
&
\centering{$\begin{pmatrix} 
\textcolor{red}{\sigma^{z}_{xx}} & \sigma^{z}_{xy} & 0 \\ 
\sigma^{z}_{yx} & \textcolor{red}{\sigma^{z}_{yy}} & 0 \\ 
0 & 0 & \textcolor{red}{\sigma^{z}_{zz}}
\end{pmatrix}$} &
\centering{$\begin{pmatrix} 
\textcolor{blue}{\sigma_{xx}} & \textcolor{red}{\sigma_{xy}} & 0 \\ 
\textcolor{red}{\sigma_{yx}} & \textcolor{blue}{\sigma_{yy}} & 0 \\ 
0 & 0 & \textcolor{blue}{\sigma_{zz}}
\end{pmatrix}$}

\\
\hline
\hline
\end{longtblr}
\end{table*}



We now turn to the spin Hall effect, evaluated within the framework of fully relativistic calculations. In the nonmagnetic phase, defined by SG 182, symmetry permits only the conventional spin Hall conductivity tensor components $\sigma^k_{ij}$, where $j$, $i$, and $k$, denoting charge current, spin current, and spin polarization of the latter, are mutually orthogonal.\cite{Roy2022_unconventional} Among these, three components are symmetry-inequivalent and thus independent. In contrast, magnetic ordering lowers the system’s symmetry due to the real-space arrangement of magnetic moments, allowing additional components of the SHC tensor. Based on the magnetic point group (MPG) symmetries, the allowed components, distinguishing between $\mathcal{T}$-even (black) and $\mathcal{T}$-odd (red), are listed in Table I.

\begin{figure*}
    \centering
    \includegraphics[width=\linewidth]{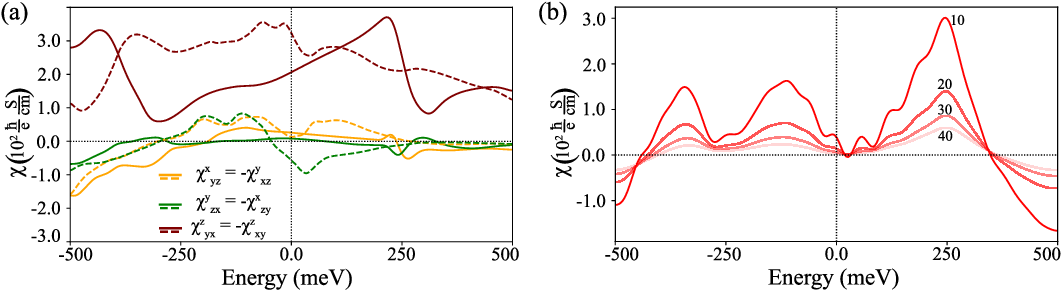}
    \caption{Relativistic calculations of the spin Hall effect in NiTa$_3$S$_6$. (a) All $\mathcal{T}$-even spin Hall conductivity tensor components calculated for nonmagnetic (dashed lines) and altermagnetic phase with N\'eel vector along the \textit{z-}direction (solid lines). (b) $\mathcal{T}$-odd spin Hall conductivity $\sigma^{x}_{xx}$ calculated for altermagnetic phase, assuming different values of $\Gamma$ = 10, 20, 30 and 40 meV. As shown in Table I, only one independent $\mathcal{T}$-odd component is present.}
    \label{fig:rel_shc}
\end{figure*}

The spin Hall conductivities of NiTa$_3$S$_6$ calculated as a function of chemical potential are shown in Fig. \ref{fig:rel_shc}. For the $\mathcal{T}$-even components, obtained using Eq. (3), we present both nonmagnetic and altermagnetic results, denoted by dashed and solid lines, respectively (Fig. \ref{fig:rel_shc}a). Overall, the magnitudes of the $\mathcal{T}$-even SHC are large, comparable with Dirac nodal line semimetals,\cite{sun2017dirac} albeit well below typical metals.\cite{zhang2021npj} In the altermagnetic phase, the $\mathcal{T}$-even SHE components are all slightly suppressed across the entire considered chemical potential range. In addition, this configuration allows four $\mathcal{T}$-odd SHC components (see Table I), which are equal by symmetry and correspond to differences in conductivity of carriers spin polarized along $+x$ and $-x$ directions.\cite{oppeneer1} These contributions, shown in Fig. \ref{fig:rel_shc}b, are generally small in magnitude, at least an order of magnitude lower than the $\mathcal{T}$-even terms at the Fermi level. 

The nonmagnetic calculations of $\mathcal{T}$-even SHC in NiNb$_3$S$_6$, shown in Fig. S16, yield slightly lower magnitudes than those in nonmagnetic NiTa$_3$S$_6$. Here, altermagnetism does not significantly suppress the overall response but introduces anisotropy among all six components, consistent with symmetry predictions (Table I). The magnetic phase also exhibits several additional $\mathcal{T}$-odd components, presented in Fig. S17. While most remain relatively small, the $\chi^{z}_{yy}$ and $\chi^{z}_{xx}$ components reach magnitudes comparable to the nonmagnetic $\mathcal{T}$-even spin Hall response. Again, the diversity in the spin Hall tensor, similarly to REE, can serve as a distinct fingerprint characterizing the in-plane N{\'e}el vector orientation, potentially enabling its experimental detection.




\section{Conclusion}
In summary, we have performed a first-principles study of the spin-orbit-related phenomena in the chiral layered materials NiTa$_3$S$_6$ and NiNb$_3$S$_6$, exploring both nonmagnetic and altermagnetic phases. In the nonmagnetic phase, both compounds exhibit persistent spin textures spanning nearly the entire Fermi surface -- a rare and desirable feature in bulk metals. This behavior leads to a sizable collinear Rashba–Edelstein effect, indicating the potential for efficient charge-to-spin conversion and long spin lifetimes. We also find a moderately large spin Hall effect in the nonmagnetic phase of both materials. 

In the antiferromagnetic regime, NiTa$_3$S$_6$ realizes an altermagnetic state with an out-of-plane N{\'e}el vector and no weak ferromagnetism, representing an unusual case of fully collinear chiral altermagnet. In NiNb$_3$S$_6$, the N{\'e}el vector lies in-plane, as confirmed by magnetic anisotropy calculations and consistent with experiments. In these cases, altermagnetic effects tend to dominate over PST, particularly in NiNb$_3$S$_6$. We find that altermagnetism generally suppresses the REE due to the competing nature of PST and altermagnetic spin textures, with a stronger reduction in NiNb$_3$S$_6$ where the N{\'e}el vector is in-plane. The emergence of additional $\mathcal{T}$-odd components of REE and SHE tensors and their larger anisotropy for the in-plane N{\'e}el vector, offers a potential way to probe magnetic order via spin responses.


\section*{Authors contributions}
K.T. performed first-principles calculations and implemented $\mathcal{T}$-even and $\mathcal{T}$-odd charge-to-spin conversion effects in PAOFLOW with the help of M.C. and M.B.N. B.K. conducted the group theory analysis. R.M.S. initialized the project. Z.H. and M.B.N. developed and implemented the VASP-PAOFLOW interface. All the authors contributed to the data analysis and discussions. C. A. and J.S. supervised the project.
\noindent

\begin{acknowledgments}
B.K. acknowledges fruitful discussions with Maxim Mostovoy and Josse Muller. J.S. acknowledges the Rosalind Franklin Fellowship from the University of Groningen. J.S. and C.C.Y. acknowledge the research program “Materials for the Quantum Age” (QuMat) for financial support. This program (registration number 024.005.006) is part of the Gravitation program financed by the Dutch Ministry of Education, Culture and Science (OCW). J.S. and B.K. acknowledge the Dutch Research Council (NWO) grant OCENW.M.22.063. C.A. acknowledges the Foundation for Polish Science project “MagTop” no. FENG.02.01-IP.05-0028/23 co-financed by the European Union from the funds of Priority 2 of the European Funds for a Smart Economy Program 2021–2027 (FENG). C.A. acknowledges support from PNRR MUR project PE0000023-NQSTI. The calculations were carried out on the Dutch national e-infrastructure with the support of SURF Cooperative (EINF-10786) and on the H\'{a}br\'{o}k high-performance computing cluster of the University of Groningen.\\
\end{acknowledgments}

\section*{Competing interests}
The authors declare no competing interests. 

\section*{Data availability}
The data that support the findings of this study will be openly available at DataverseNL.

\appendix

\section{Computational details} 
We performed DFT calculations using the Vienna Ab initio Simulation Package (VASP).\cite{vasp1,vasp2,vasp3} The exchange-correlation effects were treated within the generalized gradient approximation (GGA) using the Perdew-Burke-Ernzerhof (PBE) functional.\cite{pbe_gga} The plane-wave basis set was truncated at an energy cutoff of 350 eV, and the total energy was converged to 10$^{-7}$ eV. Brillouin zone (BZ) integrations employed a Monkhorst-Pack $k$-point mesh of $16\times16\times8$, together with Gaussian smearing of 0.05 eV.\cite{Monkhorst1976} To obtain the altermagnetic band splitting in NiNb$_3$S$_6$, we applied an effective Hubbard $U$ correction of 1.0 eV to the Ni $d$-orbitals. No Hubbard correction was used for NiTa$_3$S$_6$. Relativistic calculations were performed with SOC included self-consistently at the DFT level. We adopted a hexagonal unit cell with experimentally determined lattice constants: $a = b = 5.77$ \AA, $c = 12.03$ \AA\ for NiTa$_3$S$_6$, and $a = b = 5.76$ \AA, $c = 11.90$ \AA\ for NiNb$_3$S$_6$. The unit cells consisted of two Ni atoms, six Ta/Nb atoms, and twelve S atoms. Ionic positions were relaxed until the residual forces fell below 10$^{-3}$ eV/\AA.

As a post-processing step, we projected the DFT wave functions onto pseudo-atomic orbitals (PAOs) to construct tight-binding Hamiltonians. The following PAO basis sets were used: Ni [4s, 4p, 3d]; Ta [6s, 5p, 6p, 5d]; Nb [5s, 3p, 4p, 5p, 3d, 4d, 4f]; and S [3s, 3p, 3d]. This projection accurately reproduces the DFT band structure within an energy window of –15.0 eV to 2.8 eV for NiTa$_3$S$_6$ and –33.0 eV to 1.8 eV for NiNb$_3$S$_6$, both referenced to the Fermi level. The construction of PAO Hamiltonians and subsequent calculations were performed using our open-source Python package PAOFLOW.\cite{PAOFLOW1,PAOFLOW2} The interface between VASP and PAOFLOW was only recently developed and is utilized here. For the electronic transport calculations, the PAO Hamiltonians were interpolated onto a dense $k$-mesh of $60\times60\times40$. We used the adaptive smearing technique to smooth the single and double band integrals in all formulas.\cite{yates2007spectral}
All the calculations of charge-to-spin conversion were additionally verified by using Wannier90 package, where we implemented the same equations.\cite{pizzi2020}

Symmetry-restricted shapes of the linear response tensors are governed by the magnetic point group of each material. NiTa$_3$S$_6$ is described by MPG $6^{\prime}22^{\prime}$, and NiNb$_3$S$_6$ by $2^{\prime}2^{\prime}2$. The allowed SHC and REE tensors are calculated using the MTENSOR tool in Bilbao Crystallographic Server using Jahn symbols for $\mathcal{T}$-even REE - eV2, $\mathcal{T}$-odd  REE - aeV2, $\mathcal{T}$-even SHC - eV3, $\mathcal{T}$-odd SHC - aeV3.\cite{BCS1, BCS2, gallego2019automatic} The results are shown in Table I.

\end{document}